\documentstyle[12pt,aasms4]{article}
\font\qub=cmr10 scaled 1000
\begin{document}
\title{SBS~0335--052W -- A NEW, EXTREMELY LOW METALLICITY DWARF GALAXY
{\footnote[1]{Spectral observations reported here were obtained at the
Multiple Mirror Telescope Observatory, a joint facility of the Smithsonian
Institution and the University of Arizona}}}
\author{Valentin A. Lipovetsky{\footnote[2]{Visiting astronomer, 
National Optical Astronomical Observatories}}}
\affil{Special Astrophysical Observatory, Russian Academy of Sciences,
Nizhny Arkhyz, Karachai-Circessia 357147, Russia \\ Electronic mail: val@sao.ru}
\and
\author{Frederic H. Chaffee{\footnote[3]{Current Address: W.M.Keck
Observatory, 65-1120 Mamalahoa Hwy, Kamuela, HI 96743}} and Craig B. Foltz}
\affil{Multiple Mirror Telescope Observatory, University of Arizona, 
Tucson, AZ 85721 \\ Electronic mail: fchaffee@keck.hawaii.edu, cfoltz@as.arizona.edu}
\and
\author{Yuri I. Izotov{\footnotemark[2]}}
\affil{Main Astronomical Observatory, Ukrainian National Academy of Sciences,
Goloseevo, Kiev 252650, Ukraine \\ Electronic mail: izotov@mao.gluk.apc.org}
\and
\author{Alexei Y. Kniazev{\footnotemark[2]}}
\affil{Special Astrophysical Observatory, Russian Academy of Sciences, 
Nizhny Arkhyz, Karachai-Circessia 357147, Russia \\ Electronic mail: akn@sao.ru}
\and
\author{Ulrich Hopp{\footnote[4]{Visiting astronomer,
  Calar Alto German-Spain Observatory}}}
\affil{Universit\"{a}ts-Sternwarte, Scheiner Str. 1, D-81679, Munich, Germany\\ 
Electronic mail:  hopp@usm.uni-muenchen.de}

\begin{abstract}

We present Multiple Mirror Telescope (MMT) spectrophotometry and 3.5m Calar
Alto telescope $R$, $I$ photometry of the western component of the extremely
low-metallicity blue compact galaxy SBS 0335--052. We argue that both 
components, separated by 24 kpc, are members of an unique, physically connected
system. It is shown that SBS 0335--052W has the same redshift
as SBS 0335--052 and has
an extremely low oxygen abundance, 12 + log (O/H) = 7.45$\pm$0.22, comparable to
those in I Zw 18 and SBS 0335--052. The $(R-I)$ color profiles are very blue in
both galaxies due to the combined effect of ionized gas and a young stellar
population. We argue that SBS 0335--052W is likely to be a nearby young dwarf 
galaxy with age not exceeding 10$^8$ yr. The implications of this conclusion
to the question of galaxy formation and to the properties of primeval dwarf
galaxies are discussed.

\end{abstract}

\keywords{galaxies: abundances --- galaxies: irregular --- 
galaxies: photometry --- galaxies: evolution --- galaxies: formation
--- galaxies: ISM --- HII regions --- ISM: abundances}

\section {Introduction}

    Since its discovery as an extremely low-metallicity galaxy, the blue compact
galaxy (BCG) SBS 0335--052 (SBS -- Second Byurakan Survey)  has been identified
as a likely nearby young dwarf galaxy (Izotov et al. 1990).
This chemically-unevolved galaxy, with an oxygen abundance 1/40 solar 
(Terlevich et al. 1992; Melnick, Heydary-Malayeri \& Leisy 1992; Izotov et al.
1996), is the second most metal-deficient BCG known, after I Zw 18, whose
oxygen abundance is 1/50 (O/H)$_\odot$ (Skillman \& Kennicutt 1993). 
Further evidence in favor of the evolutionary youth of SBS 0335--052 has been presented
in several subsequent studies: Hubble Space Telescope (HST) $V, I$ imaging of 
this galaxy (Thuan, Izotov \& Lipovetsky 1996, hereafter TIL96) has shown
blue $(V-I)$ colors not only in
the region of current star formation, but also in the extended,
low-intensity envelope  4 kpc in diameter.
The emission from the underlying component, with $(V-I)$ = 0.0--0.2, appears
likely to arise from the combined effect of emission from ionized gas and 
from young ($\leq$10$^8$ yr) stars (Izotov et al. 1996).
The Very Large Array (VLA) map of SBS~0335--052 shows the presence
of a large extended HI cloud, 64 kpc in size, at the same redshift, having a mass
two orders of magnitude larger than that of the observed stars (Thuan et al. 1996).
Two prominent, slightly-resolved HI peaks, separated
by 24 kpc{\footnote[5]{Hubble constant H$_\circ$=75 km s$^{-1}$ Mpc$^{-1}$ is used
throughout the text.}}, have been detected, the eastern peak coinciding approximately with SBS 0335--052.  By comparing the VLA map to the
Digital Sky Survey (DSS), Pustilnik et al. (1996) found an optical counterpart of the western HI peak which they designated as
SBS~0335--052W. The redshift of this faint, slightly elongated compact
galaxy, measured from an 6m telescope optical spectrum, is close to that of
SBS~0335--052, suggesting that these two galaxies and the HI cloud form a single
system. Pustilnik et al. (1996) did not detect either [NII] or [SII] lines
in SBS 0335--052W and concluded that this object may
be a low-metallicity young galaxy.
However, this conclusion is based on a low S/N spectrum in the range
$\lambda$$\lambda$4800--7200\AA\, so that the oxygen
abundance was not derived. In this paper we present
new MMT spectrophotometric observations
and 3.5m Calar Alto telescope CCD $R, I$ photometry,
derive for the first time the oxygen abundance and obtain $R, I$ and $(R-I)$ 
profiles in SBS 0335--052W. These data are used to study its evolutionary status.

\section {OBSERVATIONS}
\subsection {Optical spectroscopy}
The optical spectra of SBS 0335--052W were obtained on 1996
January 19 with the Red Channel of the MMT Spectrograph, using the 300 g/mm
grating which provides a dispersion of 3 \AA\ pixel$^{-1}$ and a spectral resolution 
of about 10 \AA\ in first order. To avoid second-order contamination an L-38
blocking filter was used, and the total spectral range covered was 
$\lambda$$\lambda$3700--7300 \AA.  We used a 1$''$$ \times$ 180$''$ slit, and
the spectra were rebinned by a factor of 2 along 
the spatial axis, yielding a spatial sampling of 0.6 arcsec pixel$^{-1}$.
The total exposure time was 60 min, broken to three exposures of 20 min
for effective cosmic ray removal. The slit was oriented north-south. 
Standard stars PG 0205+134 and Feige 15 were observed for absolute flux 
calibration. Spectra of He--Ne--Ar comparison lamps were obtained
before and after each observation to provide wavelength calibration.
   
The data reduction was carried out at the NOAO headquarters in Tucson using
the IRAF\footnote[6]{IRAF: the Image Reduction and Analysis Facility is
distributed by the National Optical Astronomy Observatories, which is
operated by the Association of Universities for Research in Astronomy,
In. (AURA) under cooperative argeement with the National Science
Foundation (NSF).} software package. Procedures included bias subtraction,
cosmic-ray removal and flat-field correction using exposures of a
quartz incandescent lamp. After wavelength mapping and night sky background
subtraction each frame was corrected for atmospheric extinction and flux
calibrated. The spectrum of the brightest 1$''$$\times$2$''$ region of 
SBS 0335--052W is shown in Figure 1.  

Table 1 lists the observed line intensities and intensities corrected for interstellar
extinction for SBS 0335--052W along with the equivalent
width, EW(H$\beta$), the observed H$\beta$ emission line flux, and the extinction
coefficient C(H$\beta$). To correct for extinction we used the Galactic reddening law
 by Whitford (1958).

To derive the oxygen abundance in
SBS 0335--052W we adopted a two-zone photoionized HII region model (Stasinska
1990). The electron temperature, $T_e$(OIII), is derived from the
[OIII]$\lambda$4363/($\lambda$4959 + 5007) line intensity ratio, and the 
electron
temperature $T_e$(OII) from the relation between $T_e$(OII) and $T_e$(OIII)
fitted by Izotov, Thuan \& Lipovetsky (1994, hereafter ITL94)
for photoionized HII models
by Stasinska (1990). Because of the weakness of [OIII]$\lambda$4363, which is
barely detected, the inferred value of the electron temperature is not very reliable.
Derived values of $T_e$(OIII) and $T_e$(OII) are shown in Table 2; they are 4000 -- 5000K 
lower than those for SBS 0335--052 (Izotov et al. 1996).
For the electron number density $N_e$, we adopted 100 cm$^{-3}$ since the
[SII]$\lambda$6717, 6731 lines are weak. 
The ionic and total oxygen abundances are derived as described by ITL94 and are 
shown in Table 2. The abundances of other elements 
have not been derived because the weakness of their emission lines renders
such a determination unreliable.

\subsection {CCD photometry}

  The $R$,$I$ photometric observations of SBS~0335--052W were obtained 
during the study of SBS~0335--052 with the 3.5m telescope in
Calar Alto on 1991 October 6.
The CCD camera was installed at prime focus where, at f/3.5,
with the 1152 $\times$ 770 pixel GEC CCD (22.5 $\mu$m pixels), the field
of view is 7.3$' \times$4.9$'$.
Two exposures of 600s and 900s were taken in $R$ and $I$ respectively. Two clusters,
M92 and NGC 7790 were observed as photometric standards. The photometric
calibration was made using the data of Christian et al. (1985).

All data reduction was done with MIDAS
\footnote[7]{MIDAS is an acronym for the European Southern Observatory package
-- Munich Image Data Analysis System.}. The frames were corrected  for bias, 
dark, flat field and, in the $I$ passband, for fringe-pattern.
Unfortunately, the fringe-pattern correction was not very good and there was
some residual structure in both the horizontal and vertical directions as result of
additional noise in the CCD readout. This structure was corrected by applying
median filters separately in the X- and Y-directions.

For photometry of extended objects, we used the package SURFPHOT as well as dedicated
software for adaptive filtering developed at the Astrophysical Institute
of Potsdam (Lorenz et al. 1993).
The adaptive filter allows one to reduce pixel noise by a factor of 3--4 times without loss
of spatial resolution of the bright cores of stars and galaxies and to retain
the object's total flux as well. The smoothing scale of the adaptive filter was
11$\times$11 pixels.
Before applying an adaptive filter a special mask was built,
where all bright stars and galaxies were masked out. Such a mask is
necessary for proper determination of noise statistics used by
the filter.

The brightness profile was fitted by several different models -- 
exponential disk,
$r^{1/4}$, and linear compositions  there of.
Elliptical fitting was performed with FIT/ELL3 in the SURFPHOT 
package; the fitting algorithm
is described by Bender \& Moellenhoff (1987).
To construct a surface brightness profile we used the
geometrical average, $r$=$\sqrt{ab}$, as the average radius.

The $R$ image of SBS~0335--052W clearly shows evidence of a wisp in the eastern
direction which is absent on the $I$ frame. This could be an H$\alpha$ emission feature
similar to that observed in SBS~0335--052, but in the NW--SE direction.
Unfortunately, the slit in our MMT spectrum was oriented N--S and so did not cover that feature.

The $R$ images, isophotes and profiles with their decomposition 
for SBS 0335--052 and SBS 0335--052W are shown in Fig.2. Formally, the outer regions of both 
galaxies' profiles are well fitted by an exponential law, 
but, as described in the next section, we do not present the fitting parameters because of 
the strong bias introduced by the presence of gaseous emission.

\section {DISCUSSION}

     The existence of SBS 0335--052W poses several
important questions: a) How is this galaxy related to the brighter,
extremely low-metallicity galaxy SBS 0335--052? b) If this system of galaxies
is physically connected, what is its evolutionary state?  
Is it an evolved system of galaxies, or are we observing a system of
nearby young dwarf galaxies during their formation?
c) If this system of galaxies is young, what
do its properties tell us to expect in searches for primeval galaxies?

     In Table 3 we summarize the general characteristics of SBS 0335--052 and
SBS 0335--052W. Our measurement of the heliocentric radial velocity for
SBS 0335--052W is in excellent agreement with that of SBS 0335--052,
and confirms the conclusion of Pustilnik et al. (1996) that SBS 0335--052 and
SBS 0335--052W are physically related and are two sites of star formation in
the large cloud of neutral gas detected by the VLA observations of Thuan et al. (1996). 
Therefore, we suggest that this system is probably at a very early stage of its evolution.

Further evidence for common evolution of the two HII regions in the SBS 0335--052
system is provided by spectroscopic observations of SBS 0335--052W. The spectrum
of this HII region shows stronger low ionization [OII]$\lambda$3727 emission 
(Figure 1) and displays more moderate ionization as compared with the spectrum 
of the brighter companion.
The lower EW(H$\beta$) in SBS 0335--052W
implies that the present burst of star formation therein is older
by few times 10$^6$ yr than that in SBS 0335--052.
The oxygen abundance in SBS 0335--052W is 12 + log (O/H) = 7.45$\pm$0.22,
among the lowest known for dwarf emission-line
galaxies. It is slightly larger than 12 + log (O/H) = 7.33 derived by Izotov
et al. (1996) for SBS 0335--052, implying, along with the lower EW(H$\beta$), that the 
present burst of star formation in SBS 0335--052W
is 1--2 Myr older than in SBS 0335--052.

Based on the blue $(V-I)$ HST color
and from a combined analysis of the spectral  and photometric
observations,  TIL96 concluded that SBS 0335--052 is
probably a young galaxy with an age not greater than 10$^7$yr.
However, Izotov et al. (1996)
have shown that the underlying stellar emission in SBS 0335--052 is strongly
contaminated by emission from ionized gas.  Therefore photometric observations
alone are not enough to permit us to draw any definite conclusions about the
presence of an underlying
stellar component. Further spectroscopic observations are necessary to discriminate
between stellar and gaseous emission. 

      In Figure 2 we present the $R, I$ and $(R-I)$ profiles for both SBS 0335--052 and
SBS 0335--052W obtained at Calar Alto. The surface brightness
distribution in $I$ for SBS 0335--052 agrees well with that obtained
by TIL96, and the integrated $I$ magnitude derived from both observations agree to 
within 0.05 mag. The $R$ and $I$ profiles for SBS 0335--052
and SBS 0335--052W at large distances are fitted well by an exponential law.
The $(R-I)$ color in both HII regions gets redder with increasing radius;
however, for SBS 0335--052, the color is significantly, $\sim$0.3 mag, bluer.
In the central parts of the HII regions, emission from young massive stars
dominates.

What is the origin of the extended low-intensity underlying component?
Models of stellar population synthesis by Leitherer \& Heckman (1995) predict that,
for an instantaneous star-formation burst,
the $(R-I)$ color is $\sim$0.0 for log $t$ = 6.5
and $\sim$0.3--0.4 for log $t$ =8.0, where $t$ is age of the burst in yr.
The $(R-I)$ color for the
extended emission in SBS 0335--052 is bluer than that for a burst with log $t$ =
6.5, and the extended emission in SBS 0335--052W is bluer than that for a burst
with age log $t$ = 8.0. It is evident that the $(R-I)$ color in both HII regions
is significantly modified by the presence of ionized gas emission. This effect is
most pronounced in SBS 0335--052 because of its stronger lines. The
$R$ passband is more subject to the influence of gaseous emission because of the
presence of H$\alpha$ within the passband. 

Izotov et al. (1996) discussed the influence of ionized gas
emission on $V, R, I$ magnitudes in SBS 0335--052 and found that 2/3 of the
continuum light in the extended low-intensity component comes from a stellar population
with age 10$^8$yr. In the central part of SBS 0335--052W, EW(H$\alpha$)$\sim$500\AA\ 
and $(R-I)$ = --0.3 mag. Taking a 1500\AA\ passband 
for R, we derive $(R-I)$ $\sim$ 0.0 for the central part
of SBS 0335--052W, a color commensurate with that predicted by models of a young burst. The correction for H$\alpha$ emission in SBS 0335--052 is larger because of its
larger EW(H$\alpha$)$\sim$1000\AA\ (Izotov et al. 1996) and is equal to
$\sim$0.5 mag, just the value needed to adjust the $(R-I)$ color to that for a
young burst.  H$\alpha$ emission is not observed at distances
$r$$>$3$''$ from 
the center of SBS~0335--052W; therefore correction for gaseous emission is not
needed. 

The observed color $(R-I)$ = 0.3--0.4 in the outer envelope could be
explained by the presence of an older stellar population with age $\leq$100 Myr.
Hence, $R, I$ photometry for both SBS~0335--052
and SBS~0335--052W provides evidence of a low age and suggests a common origin 
of the star-forming regions during the last $\leq$100 Myr. The most likely mechanism for 
synchronization of
star formation in two regions separated by 24 kpc could be the contraction of
a protogalaxy. Adopting from TIL96 the number density of neutral gas to be 
$\sim$0.1--1 cm$^{-3}$, we derive
a free-fall time of $\leq$100 Myr, in agreement with estimates of the age of the stellar
population. The distribution of older stars within a region $\sim$ 1 kpc in size could be
explained by random motions of order $\sim$10 km s$^{-1}$.

The system of galaxies SBS 0335--052 is similar to the dwarf galaxy associated 
with the 1225+01 large HI  cloud in Virgo  discovered by
Giovanelli \& Haynes (1989). The HI distribution of 1225+01 shows two peaks 
separated by 15 arcmin, or 45 kpc, if a distance 10 Mpc is adopted. 
Optical knots resembling HII regions coincide with one of the peaks; the other,
however, has no optical counterpart. Salzer et al. (1991) presented a detailed
optical imaging and spectroscopic study of the dwarf irregular galaxy located
at the main peak of the HI cloud, and found the entire galaxy to be very
blue. Nebular abundances derived from their spectroscopic data reveal that 
this object is relatively unevolved chemically, although its oxygen abundance of
12 + log (O/H) = 7.66 is twice that in the main component of
SBS 0335--052 system. Furthermore, Salzer et al. found that only a tiny fraction
(0.02\% -- 0.60\%) of the mass in the NE clump of HI 1225+01 is contained in stars. 
A similar value is found in the SBS 0335--052 system (Izotov et al. 1996).
However, the luminosity and star formation rate in this system is an order of magnitude larger.
On the basis of chemical evolution and color evolution models, Salzer et al. concluded that
the galaxy associated with the HI 1225+01 cloud is still undergoing  formation
and the stellar population of this galaxy is likely to be no older than roughly
1 Gyr.

If our conclusion about the youth of SBS 0335--052 is correct, we can compare
properties of this young galaxy with theoretical predictions and properties
of high-redshift galaxies which are often considered to be primeval.
In SBS 0335--052, the Ly$\alpha$ emission line is not observed (TIL96),
contrary to theoretical predictions for young galaxies 
(Charlot \& Fall 1993). The small  mass,
$\sim$ 10$^7$M$_\odot$, of the older stellar population in SBS 0335--052
suggests that the star formation rate during the first 100 Myr was only 0.1--1 M$_\odot$
yr$^{-1}$. During this period only 1\% of the gaseous mass was transformed into
stars, contrary to predictions of several models of galaxy formation (Meier 1976;
Baron \& White 1987; Lin \& Murray 1992), where a significant fraction of the gas
is converted into stars during the collapse time. Probably  the formation of galaxies,
or at least of dwarf galaxies, is a more quiescent process with moderate conversion
of gas into stars. This could explain why primeval galaxies are not detected
in deep Ly$\alpha$ images and other surveys at large redshifts (Pritchett 1994).
Several objects have been put forward as possible primeval galaxies, mainly
on the basis of their 
very high luminosity and star formation activity (Fontana et al.
1996; Pettini, Lipman \& Hunstead 1995; Yee et al. 1996). However, most of these
candidate primeval galaxies already contain a substantial amount of heavy
elements, implying previous star formation and metal-enrichment. Properties
of these objects are very different from those for SBS 0335--052 and imply
either that these galaxies are already evolved systems or that formation scenarios for
galaxies with different masses are different. In any case, we expect that a detailed
study of the SBS 0335--052 system could provide important constraints on models of 
galaxy formation
and could assist in the search for primeval galaxies at large redshifts.

\acknowledgements
V.A.L., Y.I.I. and A.Y.K. thank R.Green and the staff of NOAO for 
their kind hospitality. The authors are very grateful to Simon Pustilnik for
stimulating discussions. Partial financial support for this international 
collaboration was made possible by INTAS collaborative research grant
No 94--2285.  Y.I.I was assisted by a travel grant from the Office of International Programs
at the University of Arizona. U.H. acknowledges the support by the SFB 375 of
the Deutsche Forschungsgemeinschaft.
C.B.F. and F.H.C. acknowledge support from NSF grant AST 93-20715.


\clearpage

\figcaption[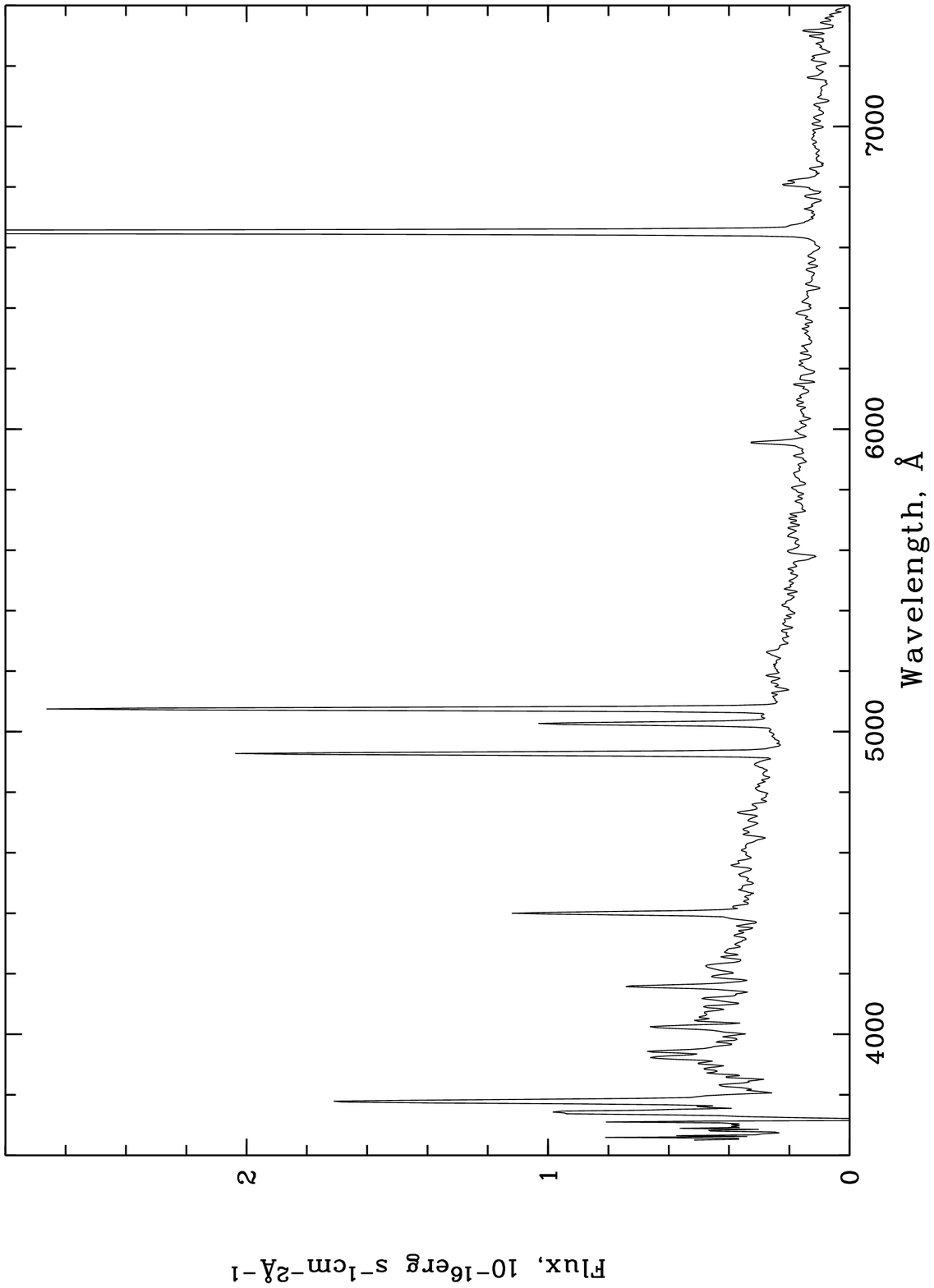]{The observed spectrum of SBS 0335--052W
obtained with Multiple Mirror Telescope.}

\figcaption[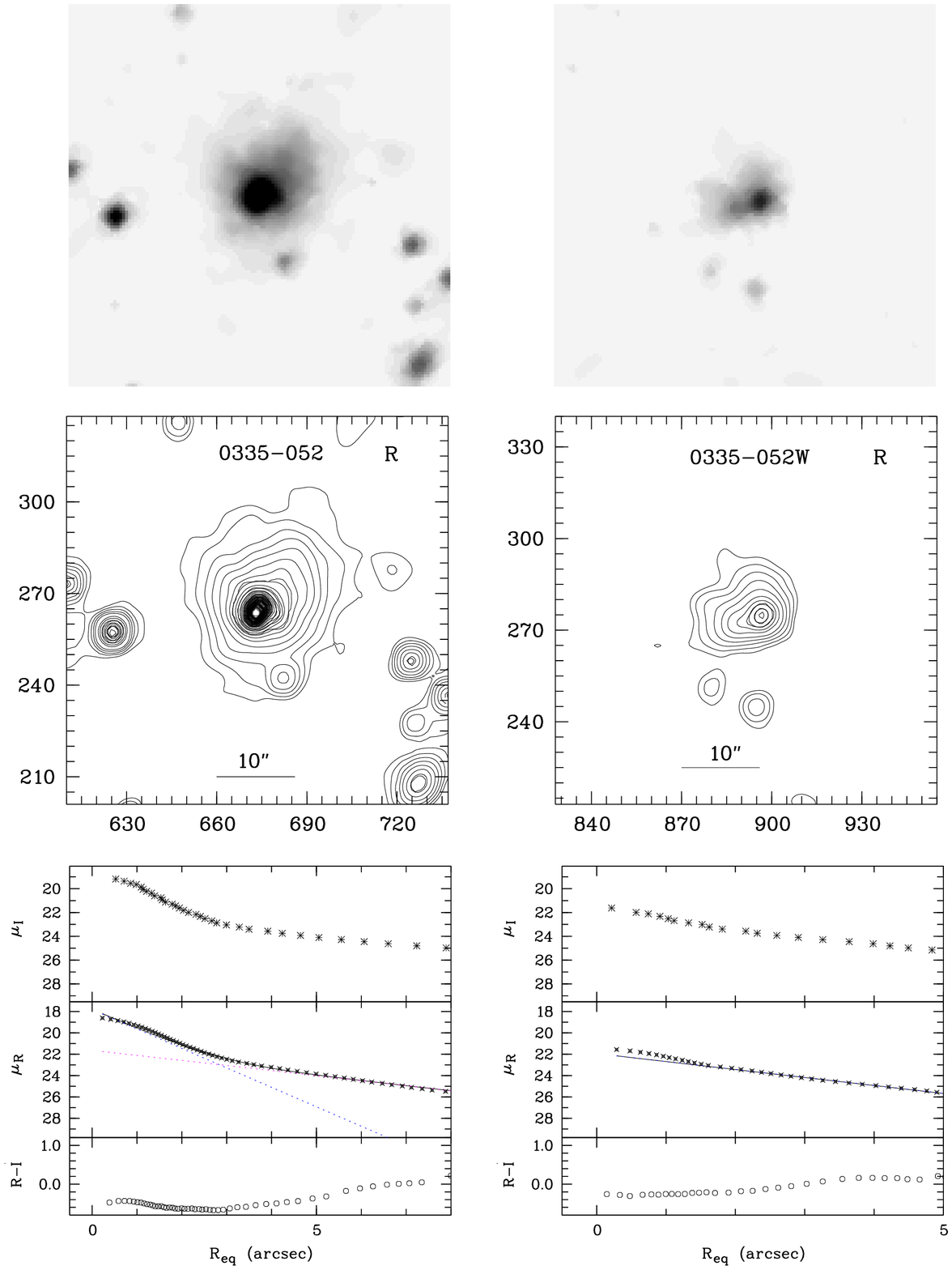]{ a) $R$-band images of SBS 0335--052 and SBS 0335--052W, obtained with
the 3.5m Calar Alto telescope. North is up, East is left;
b) Isophotes in the $R$ passband for SBS 0335--052 and SBS 0335--052W.
The outermost isophote is 26 mag arcsec$^{-2}$,
the step is 0.5 mag, the linear scale is 1$''$=263 pc;
c) The $R, I$ and $(R-I)$ profiles of SBS 0335--052 and SBS 0335--052W.
The brighter galaxy is bluer in the central region, but the colors
for both galaxies are the same in outer regions. These profiles are not compared
with those of other blue compact galaxies because the $(R-I)$ profiles are
strongly contaminated by gaseous emission in both high and low surface
brightness regions. }

\clearpage

\begin{table*}
{\small\qub
\begin{tabular}{lcc}
  Ion&F($\lambda$)/F(H$\beta$)&I($\lambda$)/I(H$\beta$)   \\ \tableline
 3727\ [O II]\ ...........................         &1.036$\pm$0.015& 1.048$\pm$0.040 \\
 3868\ [Ne III]\ ........................          &0.171$\pm$0.005& 0.172$\pm$0.005 \\
 3889\ He I + H8\ ...................              &0.160$\pm$0.005& 0.195$\pm$0.024 \\
 3968\ [Ne III] + H7\ ..............               &0.158$\pm$0.004& 0.192$\pm$0.005 \\
 4101\ H$\delta$\ ...............................  &0.202$\pm$0.004& 0.234$\pm$0.005 \\
 4340\ H$\gamma$\ ..............................   &0.460$\pm$0.005& 0.484$\pm$0.005 \\
 4363\ [OIII]\ ..........................          &0.026$\pm$0.016& 0.026$\pm$0.016 \\
 4861\ H$\beta$\ ..............................    &1.000$\pm$0.007& 1.000$\pm$0.007 \\
 4958\ [O III]\ .........................          &0.471$\pm$0.005& 0.460$\pm$0.005 \\
 5006\ [O III]\ .........................          &1.330$\pm$0.008& 1.299$\pm$0.008 \\
 5875\ He I\ ............................          &0.108$\pm$0.003& 0.104$\pm$0.003 \\
 6562\ H$\alpha$\ ..............................   &2.943$\pm$0.016& 2.795$\pm$0.015 \\
 6716\ [S II]\ ...........................         &0.067$\pm$0.003& 0.064$\pm$0.003 \\
 6730\ [S II]\ ...........................         &0.048$\pm$0.002& 0.045$\pm$0.002 \\ \\
 C(H$\beta$) dex\ ..........................&\multicolumn {2}{c}{ 0.045} \\
 F(H$\beta$)\tablenotemark{a}\ ................................. &\multicolumn {2}{c}{ 2.28} \\
 EW(H$\beta$)\ \AA\ ......................... &\multicolumn {2}{c}{ 89} \\
 EW(abs)\ \AA\ ......................... &\multicolumn {2}{c}{ 1.8} \\ 
\tablenotetext{a}{in units of 10$^{-15}$ ergs\ s$^{-1}$cm$^{-2}$}
\end{tabular}
}
\caption{Emission line intensities.}
\end{table*}

\begin{table*}

\begin{center}
\begin{tabular}{lc} 
Property&Value \\ \tableline

 $T_e$(OIII) (K)\ .............         &14,500$\pm$ 4,000   \\
 $T_e$(OII) (K)\ \,..............       &13,500$\pm$ 3,500   \\
 $N_e$(SII) (cm$^{-3}$)\ .........      & 100     \\

 O$^+$/H$^+$ ($\times$10$^5$)\ .........     &  1.27$\pm$0.95 \\
 O$^{++}$/H$^+$ ($\times$10$^5$)\ .......    &  1.57$\pm$1.14 \\
 O/H ($\times$10$^5$)\ ..............        &  2.84$\pm$1.48 \\
 12+log(O/H)\ ............                   &  7.45$\pm$0.22 \\
\end{tabular}
\end{center}
\caption{Ionic and total oxygen abundances.}
\end{table*}

\clearpage
\begin{table*}
\begin{tabular}{lcc}
 Parameter                &      0335--052W    &     0335--052 \\ \tableline

 $\alpha$\ 1950.0   .................................   &$03^h 35^m 09^s.57$&$03^h 35^m 15^s.15$    \\
 $\delta$\ 1950.0 ..................................     &$-05^\circ12'24''.0$ &$-05^\circ12'25''.9$ \\
 $R$\ mag ....................................             &   19.03          &    16.57            \\
 $R-I$\ mag ..............................            &  --0.05           &   --0.35             \\
 $R$$_{26}$ size\ kpc ............................              &   1.42            &    2.50              \\
 $V$$_{opt}$\ km s$^{-1}$ ............................     & 4069$\pm$20       & 4060$\pm$12          \\
 $V$$_{HI}$\ km s$^{-1}$   ............................  & 4006$\pm$\ 5      & 4068$\pm$\ 5          \\
\end{tabular}
\caption{Observed and derived parameters for the SBS~0335--052 system}
\end{table*}

\end{document}